\documentclass[prl,superscriptaddress,amsmath,amssymb,twocolumn]{revtex4-1}

\usepackage{lmodern}
\usepackage{graphicx}
\usepackage{mathtools}
\usepackage{amsmath}
\usepackage{amsfonts}
\usepackage{amssymb}
\usepackage{lipsum}
\usepackage[export]{adjustbox}
\usepackage{cancel}
\usepackage{dcolumn}
\usepackage{bm}
\usepackage{hyperref}
\hypersetup{linktocpage,colorlinks,citecolor={blue},pdfdisplaydoctitle=true,pdfpagemode=UseOutlines,bookmarksnumbered=true}
\usepackage{mathrsfs,dsfont}
\usepackage{color}
\usepackage{wrapfig}
\usepackage{comment}
\usepackage[normalem]{ulem}

\definecolor{cbl}{rgb}{0,0,1}
 
\definecolor{crd}{rgb}{1,0,0}

\renewcommand{\imath}{{i\mkern1mu}}

\def\qhat{\hat{q}}

\graphicspath{{Figures/}}

\def\Pc{P_0}
\def\Psat{P_{\rm sat}}

\def\nfull{m^{(\rm full)}}
\def\nch{n^{\rm ch}}
\def\nn{m}

\def\heaviside{\vartheta}

\def\tbeta{\tilde{\beta}}

\def\varmin{\chi_0}

\begin{document}

\newcommand{\titleinfo}{Darcy's law of yield stress fluids on a treelike network
}

\newcommand{\andrea}[1]{\textcolor{red}{#1}}
\newcommand{\vicio}[1]{\textcolor{blue}{#1}}
\definecolor{purple}{RGB}{150,0,96}
\newcommand{\lau}[1]{\textcolor{purple}{#1}}
\newcommand{\old}[1]{\textcolor{green}{#1}}
\title{\titleinfo}

\title{\titleinfo}
\author{Vincenzo Maria Schimmenti}
\affiliation{
Universit\'e  Paris-Saclay,  CNRS,  LPTMS,  91405,  Orsay, France
}%

\author{Federico Lanza}
\affiliation{
Universit\'e  Paris-Saclay,  CNRS,  LPTMS,  91405,  Orsay, France
}%
\affiliation{
PoreLab, Department of Physics,\\ Norwegian University of
Science and Technology, N-7491, Trondheim, Norway
}%

\author{Alex Hansen}
\affiliation{
PoreLab, Department of Physics,\\ Norwegian University of
Science and Technology, N-7491, Trondheim, Norway
}%

\author{Silvio Franz}
\affiliation{
Universit\'e  Paris-Saclay,  CNRS,  LPTMS,  91405,  Orsay, France
}%

\author{Alberto Rosso}
\affiliation{
Universit\'e  Paris-Saclay,  CNRS,  LPTMS,  91405,  Orsay, France
}%

\author{Laurent Talon}
\affiliation{
Universit\'e  Paris-Saclay,  CNRS,  FAST,  91405,  Orsay, France
}%

\author{Andrea De Luca}
\affiliation{Laboratoire de Physique Th\'eorique et Mod\'elisation,
CY Cergy Paris Universit\'e, \\
\hphantom{$^\dag$}~CNRS, F-95302 Cergy-Pontoise, France}

\begin{abstract}
Understanding the flow of yield stress fluids in porous media is a major challenge. In particular, experiments and extensive numerical simulations report a non-linear Darcy law as a function of the pressure gradient. In this letter, we consider a tree-like porous structure for which the problem of the flow can be resolved exactly thanks to a mapping with the directed polymer (DP) with disordered bond energies on the Cayley tree. Our results confirm the non-linear behavior of the flow and expresses its full pressure-dependence via the density of low-energy paths of DP restricted to vanishing overlap. These universal predictions are confirmed by extensive numerical simulations.
\end{abstract}
\date{\today}

\maketitle

In a series of experiments during the nineteenth century, Henry Darcy studied the flow of water in a cylinder filled with sand \cite{Darcy} and established the  empirical law for the flow rate $Q$ as a function of the pressure difference $P$ between the two ends of the cylinder
\begin{equation}
\label{eqn:darcynewt}
    Q= \kappa R^2 P/(\eta L)
\end{equation}
here  $R$ and $L$ are respectively the radius and the length of the cylinder, $\eta$ is the viscosity of the fluid. The permeability, $\kappa$, has the dimension of a surface and measures the ability of a given porous medium \cite{b88,s11,b17,ffh22} to transmit a fluid.  
Darcy gave an interpretation to the permeability assuming that, in a medium, the flow is possible only along non-intersecting thin channels, each of radius $~ R_c \ll R$. The flow along a single channel is given by the Poiseuille's law which holds for empty cylinders and the total flow can be written as  $Q = \pi R^2  n^{\rm ch} \pi R_c^4 P/(8 \eta L)$, with $n^{\rm ch}$  the number of channels per unit surface. Hence, the permeability can be identified as $\kappa = \pi  n^{\rm ch}  R_c^4/8$.  The network of the channels of a real porous medium is more  complex: channels have  heterogeneous shape and can intersects. However the Darcy law is  valid as far as the number of channels remains pressure-independent. This is not the case for yield stress fluids, such as suspensions \cite{fall09}, gels \cite{piau07}, heavy oil \cite{pascal81}, slurries or cement \cite{coussot05} for which a minimum yield stress , $\sigma_Y$, is needed to flow \cite{bingham22}. 
Hence, at low pressure gradient, these yield stress fluids behave like a solid and no flow is measured. However, increasing the pressure gradient, they start  flow along more and more channels. Experiments \cite{al-fariss87,chase05} and numerical simulations \cite{lopez03,balhoff04,chen05} indicated that the Darcy law is modified: below a threshold pressure $\Pc$ no flow occurs while, above it, the flow grows non-linearly with $P$. Three flowing regimes are observed  \cite{talon13b,chevalier15a} : i) initially the flow grows linearly in $P-\Pc$, but with an effective permeability which is very small  ii) for larger pressure the flow grows non-linearly as $(P-\Pc)^\beta$ (with $\beta \approx 2$)  \cite{rh87, ChenDeLucaRossoTalon}. iii) only above a saturation pressure $\Psat \gg \Pc$, the flow recovers the linear growth with the standard permeability $\kappa$ \cite{barnes00}.

\begin{figure}[ht]
 	\centering
 	\includegraphics[width=0.45\textwidth, trim={1cm 0 0 0},clip]{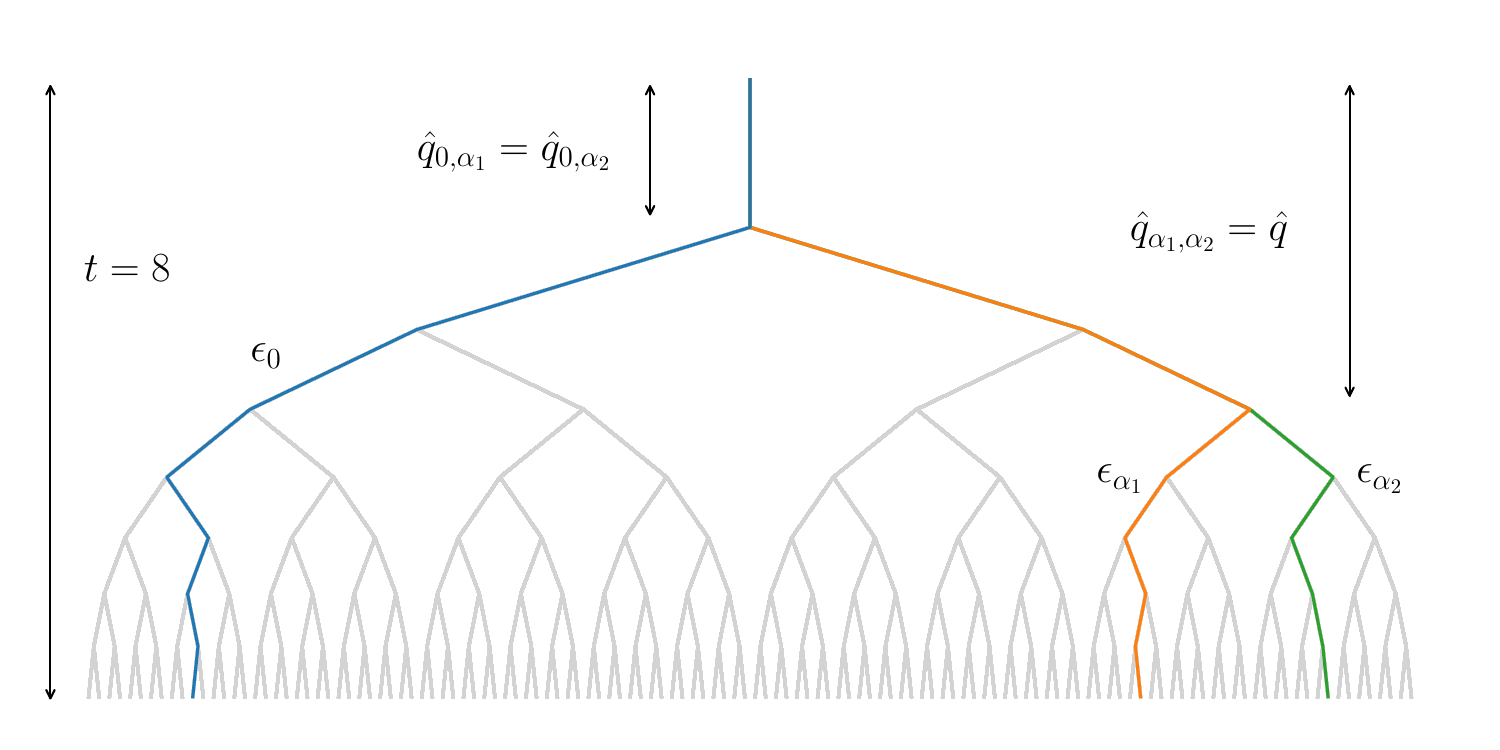}
 	\includegraphics[width=0.45\textwidth]{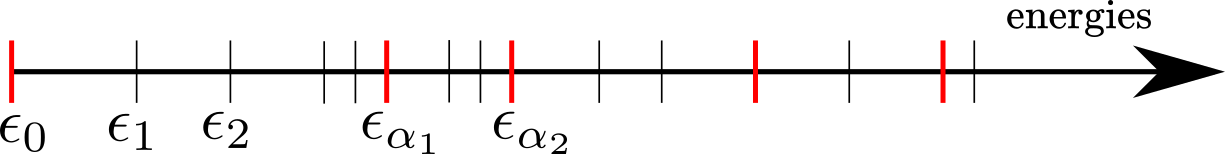}
 	\caption{
 		Top: Binary Cayley tree with $t=8$ levels. The first channels where flow occur are the leftmost path in blue, at the pressure $P_0$, the middle path in orange $\alpha_1$, at pressure $P_1>P_0$ and the rightmost path in green $\alpha_2$,  at pressure $P_2>P_1$. We denote with $\qhat_{\alpha_1 \alpha_2}$ the overlap between $\alpha_1$ and $\alpha_2$, namely the length of the common path between them (here $\qhat_{\alpha_1 \alpha_2}=3$). Similarly  $\qhat_{0, \alpha_1}$ and $\qhat_{0, \alpha_2}$ are the overlaps of $\alpha_1$ and $\alpha_2$ with the blue path
 		(here $\qhat_{0,\alpha_1}= \qhat_{0,\alpha_2}=1$) and $\qhat$ is the maximal overlap between all of them (here $\qhat=3$). Bottom: sorted energies of the associated directed polymer  $\epsilon_0 < \epsilon_1< \epsilon_2, \ldots$.  The energies corresponding to small overlap paths are in red. In the large-$t$ limit we show that $\qhat \ll t$ and $P_1=\epsilon_{\alpha_1}, P_2=\epsilon_{\alpha_2}, \ldots$.
 	}
 	\label{fig:tree}
 \end{figure}

Despite these detailed observations, a theoretical explanation for the non-linearity is still lacking. In this letter, we propose both an explanation and provide an explicit prediction for the modified Darcy law. We consider a porous structure with the geometry of a binary Cayley tree with $t$ levels (see Fig.~\ref{fig:tree}). This geometry is the simplest with intersecting channels and is realistic for several biological networks (e.g. alveoli system in lungs \cite{Mauroy2004} or leaf veins). In this work, we establish a precise mapping between the Darcy problem for yield stress fluids and the directed polymer on Cayley tree, a model displaying one-step replica symmetry breaking (1-RSB) \cite{mezard1987spin,Derrida1988}. Each individual channel in the porous structure has a pressure threshold which we identify with the energy of the directed polymer represented by that channel. We show that the first channels where flow occurs correspond to those low-energies directed polymers with small overlaps (i.e. they share a short common path). Next we modified the Kolmogorov Petrovsky Piskunov (KPP) approach proposed in \cite{Brunet2011} to determine analytically their number, $n^{\rm ch}$, as a function of $x=P-\Pc$ (see Eq. \eqref{eq:mqlim}). Finally, we determine explicitly  the low pressure behaviour, described by Eq. \eqref{eq:limq}, the linear high pressure regime of Eq. \eqref{eq:qt_final} as well as the saturation pressure $\Psat$  (see Eq. \eqref{eq:Psat}).   At this pressure, the fluid flows along $\sim t$ channels almost non-overlapping. By further increasing  the pressure, new channels open without significantly changing the permeability of the network. This predictions are confirmed by direct numerical simulations.

 \paragraph{Mapping to the directed polymer and large-$t$ limit. ---}
Our model is a Cayley tree pore network filled by a Bingham fluid. A pressure $P$ is applied on the root pore and a zero pressure at the leaves. In this model large open pores with a well-defined pressure are connected by tubes of random radius and length. The modified Poiseuille law for a Bingham fluid is an open problem, but in the limit $P \gg \tau$ takes a simple for $Q_{\rm Pois}(P) = \frac{\pi R^4}{8 \eta L}(P - \tau)_+$ (see \cite{chen05}). Here we denote $(x)_+ = \max(0,x)$ and  $\tau=L \sigma_Y /R$ and consider $P>0$.
We consider the simplified case in which only the thresholds fluctuate and the flow in a tube between the pore $i$, at the pressure $P_i$, and the pore $j$, at pressure $P_j<P_i$, reads:
\begin{equation}
    \label{eq:bingham}
    Q_{ij} = (P_i-P_j-\tau_{ij})_+
\end{equation}
The threshold $\tau_{ij}$ is a random variable, drawn from a distribution $p(\tau)$. Hence a tube is open if $P_i-P_j > \tau_{ij}$.
A pore has one incoming tube and two outgoing tubes and inside it Kirchhoff's law holds: the incoming flow must be equal to the sum of the outgoing ones. It follows that, given an open incoming tube, there must be at least one outgoing open tube. Thus along a channel from the root to a leaf flow occurs if all its $t$ tubes are open. 

As a consequence, the pressure $P_0$ at which the first channel opens is given by:
\begin{equation}
\label{eqn:pressure0}
    P_0 = \min_\alpha \sum_{(ij)\in \alpha} \tau_{ij}
\end{equation}
where $\alpha$ labels the $2^{t-1}$ directed path connecting the root to a leaf. 
As observed in \cite{ChenDeLucaRossoTalon}, the threshold pressure $P_0$ identifies with the ground state energy of the associated directed polymer (DP) model. We define $\epsilon_\alpha$ the energy of a directed path $\alpha$ as the sum of the thresholds along $\alpha$:
\begin{equation}
    \epsilon_\alpha = \sum_{(ij) \in \alpha} \tau_{ij}
\end{equation}
It follows that $P_0=\min_\alpha \epsilon_\alpha$.
The key point of this letter is that also the pressures $P_1, P_2, \dots$ at which a new channel opens are related to the low-energy levels of the DP. It is useful to label the directed paths $\alpha$ by ordering the energies as $\epsilon_0 < \epsilon_1 < \epsilon_2 < \ldots$. 

Using Kirchhoff's law (see section A of \cite{Note1}) we compute the explicit expression for $P_1$:
\begin{equation}
\label{eqn:pressure1}
    P_1 = \epsilon_0 + \min_{\alpha \neq 0} \frac{\epsilon_\alpha-\epsilon_0}{1-\qhat_{0\alpha}/t} = 
     \epsilon_0 + \frac{\epsilon_{\alpha_1}-\epsilon_0}{1-\qhat_{0\alpha_1}/t} 
\end{equation}
Here $\hat{q}_{0\alpha}$ stands for the overlap  between the $\alpha$ channel and the ground state, namely the number of common tubes between the two channels. The path $\alpha_1$ realises the minimum and it is the channel that opens just above $P_1$. 
It is crucial to remark that the minimization involves two terms: the term $\epsilon_\alpha-\epsilon_0$ favors low-energy paths while the term $1/(1-\qhat_{0\alpha} /t)$ selects the ones with a small overlap with the ground state.
The directed polymers on Cayley tree display one step replica symmetry breaking 1-RSB. This means that in the limit $t \to \infty$ the overlap among any two low-energy directed paths is either $O(1)$ or $\sim t$ \cite{mezard1987spin} (finite $t$ corrections are also known \cite{DerridaMottishaw, CaoRossoSantaDouss}).  Hence, in this large-$t$ limit, the channel $\alpha_1$  corresponds to a path with low-energy and low-overlap ($\qhat_{0\alpha_1} = O(1)$), so that $\lim_{t \to \infty} 1/(1-\qhat_{0\alpha_1} /t) = 1$ and $P_1 = \epsilon_{\alpha_1}$. Moreover, the total flow reduces to the sum of the contribution from each channel, namely $Q_t(P) = (P-P_0)/t + (P-P_1)/t$ for $P \in [P_1,P_2]$ (see section A of \footnote{See Supplemental material which includes Refs. \cite{Derrida1988,Brunet2011,kolmogorov1937investigation,Bramson83,MAJUMDAR2003161,Bramson83,Brunet2011,ChenDeLucaRossoTalon} for additional information about each of the three sections of the main paper: A) Mapping to the directed polymer and large-$t$ limit ; B) KPP approach; C) Determination of the flow. }).

The same property holds for higher pressures (see section B of \cite{Note1} ): the channels $\alpha_2, \alpha_3, \dots$ correspond to paths with the low-energy and low-overlap and $P_2, P_3, \dots$ coincide with the energies $\epsilon_{\alpha_2}, \epsilon_{\alpha_3}, \dots$ (see sorted energies of Fig. \ref{fig:tree} bottom).
Similarly the total flow reduces to the sum of the contribution from each channel.

As a consequence the computation of the flow problem reduces to determine the growth of the number of open channels $\nch_t(x)$ as a function of $x=P-P_0$. This number identifies with the number of low-energy levels  of the directed polymer, provided they have low overlap among them. We compute this number adapting the tools introduced in \cite{Brunet2011} and based on the mapping to the discrete Kolmogorov-Petrovsky-Piskunov (KPP) equation~\cite{Derrida1988, MAJUMDAR2003161,PhysRevE.62.7735,PhysRevE.65.036127, Arguin2011,Ramola2015}.
\paragraph{KPP approach. ---}
To begin, we introduce the number of energy levels with energy smaller than $\epsilon_0+x$, namely $\nfull_t(x) = \sum_{\alpha} \heaviside(x - (\epsilon_\alpha - \epsilon_0))$
($\heaviside(x)$ is the Heaviside theta function). In \cite{Brunet2011} (for a self-contained derivation see section B of \cite{Note1})
  $\overline{\nfull_t(x)}$ (the overbar stands for the average over the random thresholds) is expressed as  $ \overline{\nfull_t(x)} = \int dx' \, r_t(x';x)$. The function $r_t(x';x)$ satisfies the following recursive equation:
\begin{flalign}
\label{eqn:rtrecursive}
& r_{t+1}(x'; x) = 2\int d\tau p(\tau) \Omega_t(x'-\tau)  r_t(x'-\tau; x) \\
\label{eqn:mininvcdf}
 &   \Omega_{t+1}(x) = \int d\tau p(\tau) \Omega_t(x-\tau)^2
\end{flalign}
Here $p(\tau)$ the thresholds distribution and the initial conditions read $r_1(x';x)=p(x+x')$ and $\Omega_1(x) = \int_x^\infty d\tau p(\tau)$. 
Equation \eqref{eqn:mininvcdf} is the discrete KPP equation  and the function $\Omega_t(x)$ is the probability that the ground state energy of the DP on a Cayley tree of $t$ levels is larger than $x$. Hence equation \eqref{eqn:mininvcdf} corresponds to growing a $t+1$-level tree starting from two $t$-level trees \cite{Derrida1988}. 

\begin{figure}[ht]
\includegraphics[width=0.95\columnwidth, height=5cm]{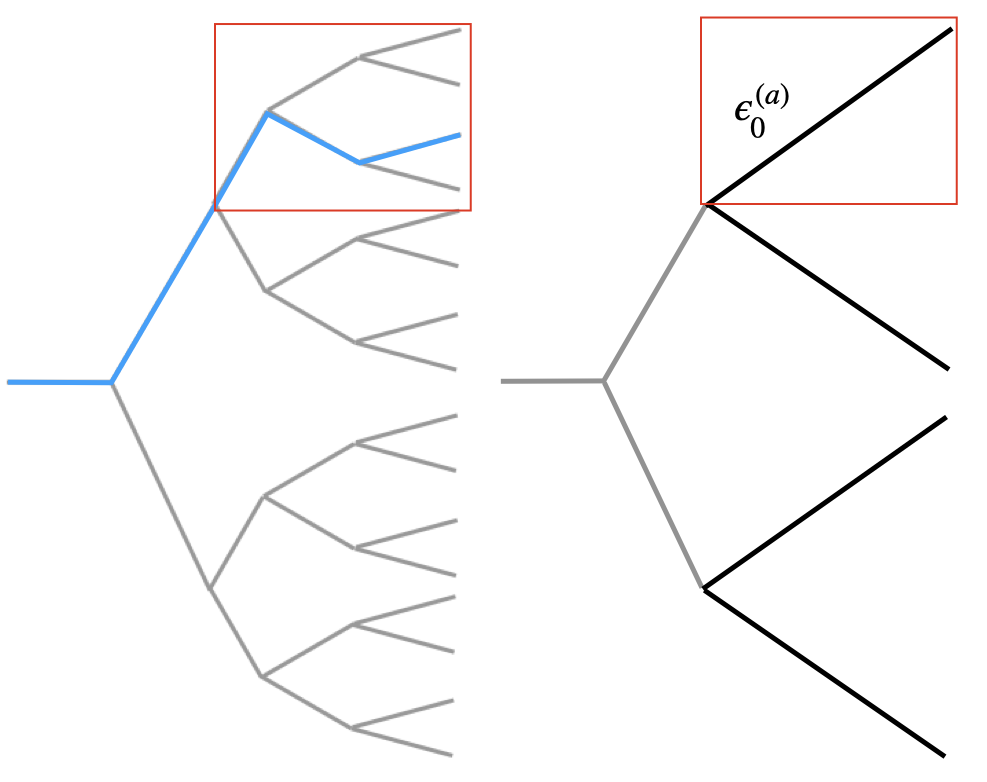}
\caption{Left: An example of Cayley tree. The minimal path of the topmost subbranch after $\qhat = 2$ generations is shown highlighted in blue. Right: Pruning of the full Cayley tree, where within each subtree starting from the $\qhat=2$ generation, only the minimal path is retained. \label{fig:bethepruning}
}
\end{figure}

\begin{figure*}[ht]
    \includegraphics[width=0.32\linewidth]{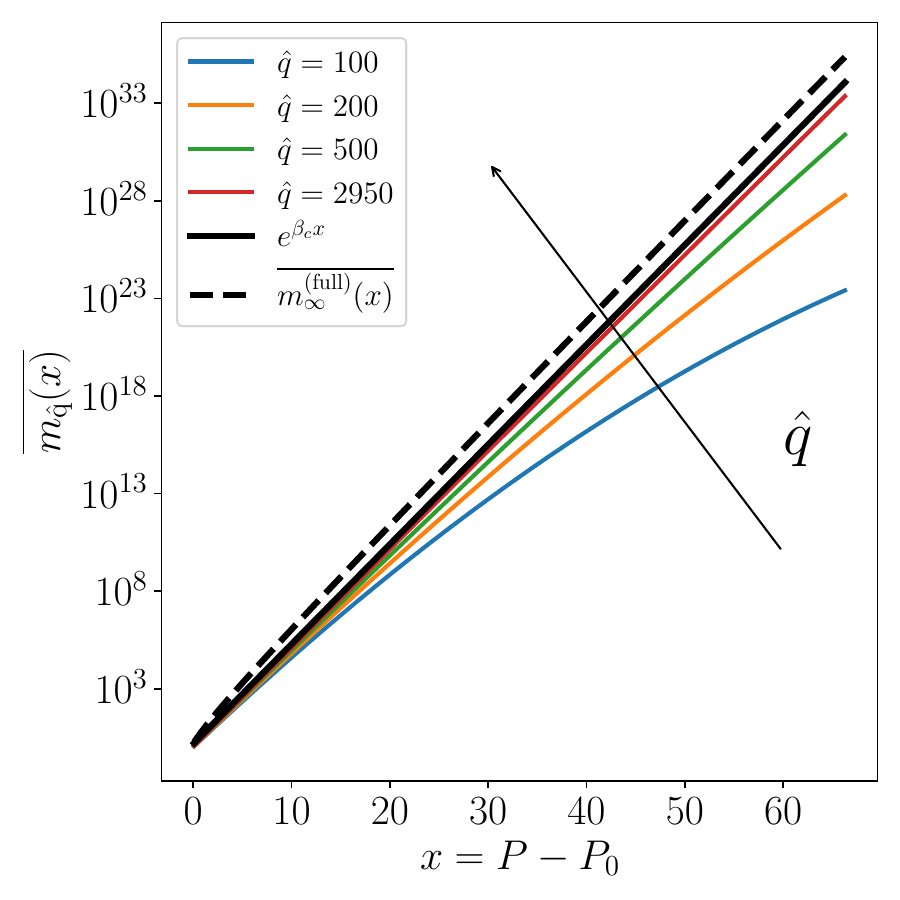} 
    \includegraphics[width=0.32\linewidth]{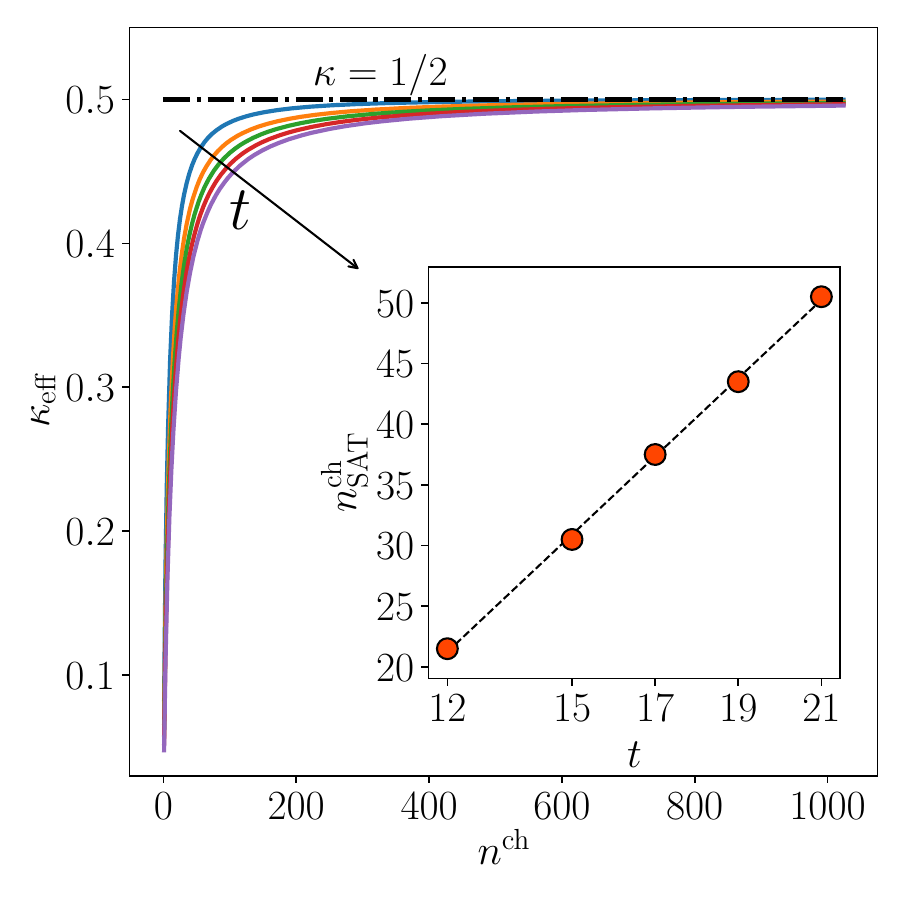}
    \includegraphics[width=0.32\linewidth]{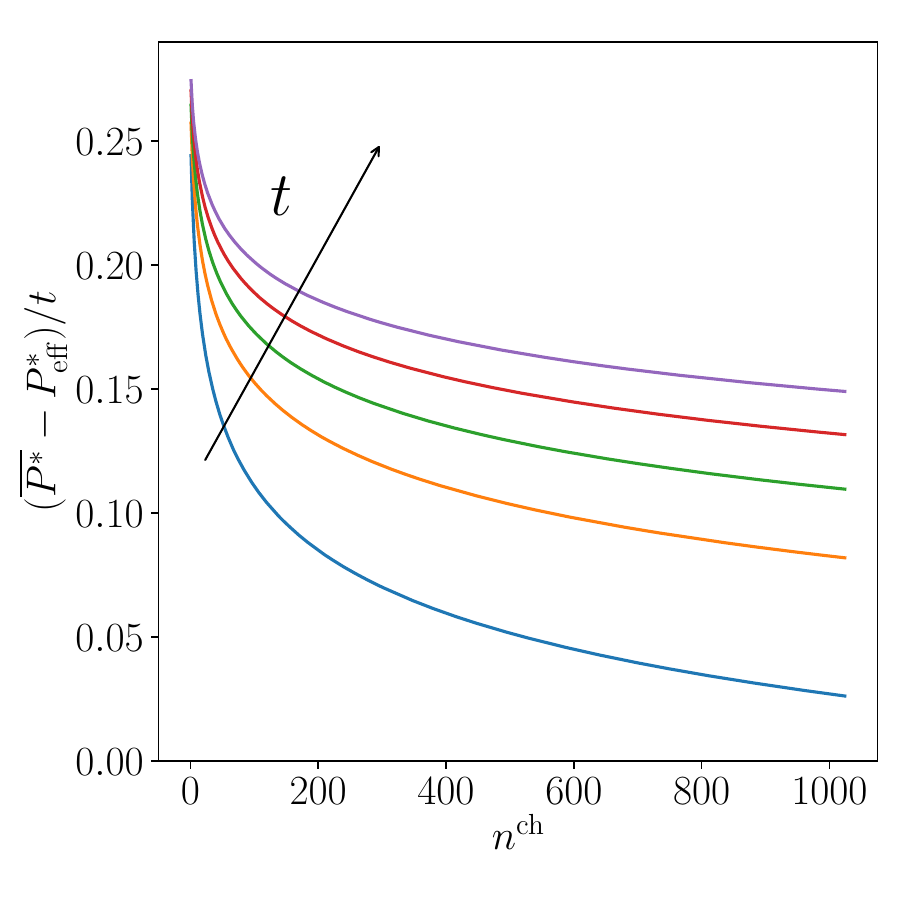}
    \caption{
    Left:  $\overline{\nn_{\qhat}(x)}$ and $\overline{\nfull_\infty (x)}$ (dashed black line)  obtained  by numerical integration of Eq.(\ref{eqn:rtrecursive}) and  Eq.(\ref{eqn:mininvcdf}) with different initial conditions. 
    Middle and Right: exact numerical simulations on finite Cayley tree of moderate moderate sizes $t=12,15,17,19,21$. The opening of the first $\sim t$ channels is sufficient to saturate the effective permeability,       $\kappa_{\rm eff}(\nch_{\rm SAT}) =0.4 =0.8 \kappa$ 
    (inset). However $P^*_{\rm eff}$ displays a much slower evolution and it is still far from saturation when  $\nch \sim 1000\gg  t$. The simulations were carried out using a Gaussian distribution $p(\tau)$ with zero mean and variance $\sigma^2=1/12$. The corresponding value for $\beta_c$ from (\ref{eqn:betac_cbeta}) is $\beta_c=\sqrt{2\ln2}/\sigma$.
    }
    \label{fig:panel}
\end{figure*}

To determine the Darcy flow we are interested in a subset of low-energy paths, namely the ones that contribute to $\nch_t(x) = \sum_i \heaviside(x - (P_i - P_0))$ with $i=1 \dots 2^{t-1}-1$.
As discussed above, in the limit $t \to \infty$, the open channels coincide with the low-energy paths with vanishing overlap and the pressures $P_1, P_2, \dots$ with the corresponding path energies $\epsilon_{\alpha_1}, \epsilon_{\alpha_2}, \dots$.
To make progress we introduce the quantity $\nn_{\qhat}(x)$ which counts, in an infinite tree, the number of paths with energy $\epsilon_\alpha \leq \Pc + x$ and maximum overlap between them $\qhat$. With this prescription, we  immediately take limit $t \to \infty$  and only consider low-energy paths with vanishing overlap, $\qhat \ll t$. This remains valid even in the limit $\qhat\to \infty$, so that
\begin{equation}
\label{eq:nnhatch}
     \nch_{\infty}(x)
    =\lim_{\qhat \to \infty} \nn_{\qhat}(x) \;.
\end{equation} 

To compute $\overline{\nn_{\qhat}(x)}$, we modify the KPP approach introduced for $\overline{\nfull_t(x)}$.
For this, we introduce a pruning procedure (see fig.~\ref{fig:bethepruning}): at the level $\qhat$ of the full Cayley tree, there are $2^{\qhat}$ subtrees labelled by $a = 1,\ldots, 2^{\qhat}$. We replace each of these subtrees with a single tube with a minimum energy $\epsilon^{(a)}_0$. This way we obtain a tree of $\qhat$ levels containing the $2^{\qhat}$ low energies paths  with maximal overlap $\qhat$.
This procedure is equivalent to growing a tree with $\qhat$ levels where the leaves thresholds are drawn from the distribution of the minimum of an infinite tree. The probability $w_{\rm min}(x)$ that the minimal energy of an infinite tree is larger than $x$ is obtained from the fixed-point travelling wave solution of equation \eqref{eqn:mininvcdf}:
\begin{equation}
\label{eqn:kppfixedpoint}
    w_{\rm min}(x+c(\beta_c)) = \int d\tau p(\tau) w_{\rm min}^2(x-\tau)
\end{equation}
where $c(\beta_c)$ is the minimal value for which \eqref{eqn:kppfixedpoint} has a solution and its value can be obtained as:
\begin{subequations}
\label{eqn:betac_cbeta}
    \begin{align}
& \beta_c = \arg \min_\beta c(\beta) \\
& c(\beta) = \frac{1}{\beta} \log \left( 2 \int d\tau p(\tau) e^{-\beta \tau} \right)
\end{align}
\end{subequations}
The function $w_{\rm min}(x)$ is a sigmoid with $w_{\rm min}(x) \simeq 1 - x \exp{(\beta_c x)}$ for $x \to -\infty$ and $w_{\rm min}(x)\simeq 0$ for $x\to \infty$. The solution of eq. \eqref{eqn:kppfixedpoint} is defined up to an arbitrary shift that we set to zero for simplicity. For large but finite $t$, it has been proven \cite{Bramson83} that :
	\begin{equation}
		\label{eqn:P0_appendix}
		P_0 = \epsilon_0 = -c(\beta_c)t + \frac{3}{2 \beta_c} \log t + \varmin,
	\end{equation}
where the first two terms are deterministic while $\varmin $ is a random variable of order $1$ distributed according to $-w'_{min}(\varmin)$.
In our problem, the limit $t \to \infty$ can be safely taken as the divergent deterministic part is unimportant, being the same for all leaves.

Thus, to compute $\overline{\nn_{\qhat} (x)}$, one has to implement the recurrence in $\qhat$ instead of $t$, replacing $r_t(x';x)$ with $r_{\qhat}(x';x)$. 
Moreover, in equation \eqref{eqn:rtrecursive}, $\Omega_t(x)$ is replaced with $w_{\rm min}(x+c(\beta_c) \qhat)$ and the initial condition is 
$r_{\qhat=1}(x';x) = -w_{\rm min}'(x+x')$.
 Finally, as before, $\overline{\nn_{\qhat} (x)} = \int dx' \, r_{\qhat}(x';x)$. 

In the limit $t\to \infty$ a closed-form expression for $\overline{\nfull_\infty(x)}$ is not known \cite{10.21468/SciPostPhys.1.2.011} however in \cite{Brunet2011} it was shown numerically that the following asymptotic holds:
\begin{equation}
    \overline{\nfull_\infty(x)} \stackrel{x \gg 1}{=} A x e^{\beta_c x}
\end{equation}
with $A$ a non-universal $O(1)$ constant.
On the contrary, direct numerical integration of $r_{\qhat}(x';x)$ (Fig. \ref{fig:panel} (right)) show that, when $\qhat \to \infty$, the expression of   $\nch_{\infty}(x)$  is:
\begin{equation}
\label{eq:mqlim}
    \nch_{\infty}(x) = \lim_{\qhat \to \infty} \overline{\nn_{\qhat}(x)} = e^{\beta_c x}.
\end{equation}
In \cite{Note1}, we also provide an analytical argument to support this result.

\paragraph{Determination of the flow. ---} At low pressure, for a fixed  $x = P -\Pc$, there is are finite number of channels, sharing low overlap and each supporting a flow $(x-x')/t$, being $\Pc+x'$ its opening pressure. In this regime the total flow reduces the sum of independent contributions $Q_t(\Pc + x) =\int_0^x dx' \,\nch_{t}(x - x')/t$. For large $t$ this leads to
\begin{equation}
\label{eq:limq}
 \overline{Q_t(\Pc + x)} = \frac{e^{\beta_c x}-1}{\beta_c t}\; \quad \textrm{with } P \gtrsim \Pc 
\end{equation}
This expression captures the first two regimes of the flow: when $P \to \Pc$ the flow is linear $Q(P) = (P-\Pc)/t$, for larger pressure a non linear regime takes over.
At very high pressure instead all channels are open and we recover the second linear behavior with: 
\begin{equation}
\label{eq:qt_final}
    Q_t(P)=\kappa(P-P^*) \quad \textrm{with } P \to \infty
\end{equation}
For large $t$, $\kappa = 1/2$ and $P^* = \overline{\tau} t$ (with $\overline{\tau}=\int \tau p(\tau) d\tau$. See section C of \cite{Note1}). 

The crossover between the non-linear regime of equation \eqref{eq:limq} and the linear regime of equation \eqref{eq:qt_final} occurs at the pressure $P_{\rm sat}$. An estimation of  $P_{\rm sat}$ is obtained by matching the effective permeability at low pressure,  $\overline{\kappa_{\rm eff}} = d\overline{Q_t}/dP \sim e^{\beta_c (P - \Pc)} / t$,  with the value $\kappa=1/2$ at high pressure:
\begin{equation}
\label{eq:Psat}
    P_{\rm sat} =\Pc + (1/\beta_c) \ln t
\end{equation}
As consequence, at the saturation pressure, the number of channels obtained by eq. \eqref{eq:mqlim} is $\sim t$. Let us comment about this result.
When the pressure is slightly above the minimal value $\Pc$, only a single channel is open and $\kappa_{\rm eff}$ is $\sim 1/t$. Increasing the pressure slightly more ($\sim \ln t$), it  is enough to have $\sim t$ channels with very small overlap between them and to reach the total permeability $\kappa$. Note that this number is very small compare to $2^{t-1}$, the total number of directed  paths.   At even larger pressure, the fluid flows indeed in more and more channel, but this does not affect  much the permeability of the network. To check these results we carried out exact numerical simulations on the Cayley tree with moderate $t$. For a given finite tree the flow curve as function of the pressure is a piece-wise linear function, with breakpoints at $P_0, P_1,\ldots, P_{\nch{}}$, the pressures at which a new channel opens. For each segment the flow reads $Q(P)= \kappa_{\rm eff} (P- P_{\rm eff}^*).$ The first parameter, $\kappa_{\rm eff}$, is the permeability of the set of open channels, while $P_{\rm eff}^*$ depends on the threshold along them.  In Fig. \ref{fig:panel}, we study their behavior as a function of $\nch{}$. In Fig.   \ref{fig:panel} middle, we observe that the permeability grows quickly with $\nch{}$ and after  $n_{\rm SAT}^{\rm ch}\sim t$ reach the value  $\kappa=1/2$ (see inset of Fig.  \ref{fig:panel}middle). The converse is not true for $P_{\rm eff}^*$ which grows slowly as shown in Fig. \ref{fig:panel} right.
\paragraph{Conclusions. ---}

In this work, we show that the Darcy problem with a yield stress fluid  is closely related to the associated directed polymer. In particular, in the limit of large trees, a direct mapping emerges between $n^{\rm ch}$ and low-energy directed  paths with zero-overlap. Thanks to this identification, we derive a simple universal expression for the flow as a function of the applied pressure. Equation \eqref{eq:limq} is independent of most microscopic details. However,  the threshold distribution and the tree branching ratio set the parameter $\beta_c$. \\ 
The next big challenge would be to solve the problem of the flow in finite dimension. In particular, it would be interesting to understand the  role  of the low-energy and  low-overlap energy levels.  Those low-overlap excitations are abundant in mean-field glassy disordered systems but their number is suppressed in finite dimension.
For this reason, their role in realistic set-ups has always been controversial. However, in the Darcy problem, excitations with high overlap are strongly penalized and they are inessential in increasing the flow, independently of the spatial dimension. For this reason,  the Cayley tree solution of the flow can give important insights on finite dimensional porous media. In particular the high pressure behaviour of Eq.(\ref{eq:qt_final}) holds in all dimensions and we expect that the scenario depicted in Fig.\ref{fig:panel} middle and right holds as well. In a real porous medium,  we predict that the effective permeability grows initially fast and saturates to $\kappa$, while $P^*_{eff}$ evolves slowly to $P^* = t \bar \tau$.
Moreover, from the Cayley tree solution, we know that the permeability of the network of flowing channels is governed by a small number of independent channels.
It would be interesting to understand if this remains true in finite dimensions.

\paragraph{Acknowledgements. ---}
We thank X. Cao and V. Ros for discussions and suggestions. ADL acknowledges support by the ANR JCJC grant ANR-21-CE47-0003 (TamEnt). We acknowledge the support of the Simons foundation (grant No. 454941, S. Franz).
This work was partly supported by the Research Council of Norway through its Center of Excellence funding scheme, project number 262644. Further support, also from the Research Council of Norway, was provided through its INTPART program, project number 309139. This work was also supported by "Investissements d'Avenir du LabEx" PALM (ANR-10-LABX-0039-PALM).

\let\oldaddcontentsline\addcontentsline
\renewcommand{\addcontentsline}[3]{}
\bibliographystyle{apsrev4-1}
\bibliography{biblio}
\let\addcontentsline\oldaddcontentsline

\onecolumngrid

\newpage 
\setcounter{equation}{0}
\setcounter{figure}{0}
\makeatletter
\renewcommand{\theequation}{S\arabic{equation}}
\renewcommand{\thefigure}{S\arabic{figure}}
\renewcommand{\bibnumfmt}[1]{[S#1]}
\renewcommand{\citenumfont}[1]{S#1}
	\begin{center}
		{\Large Supplementary Material \\
			Darcy's law of yield stress fluids on a treelike network
		}
	\end{center}

\section{A. Mapping to DP and large-$t$ limit}
In this section we report the derivation of the expression for $P_1$ used in the main text and $P_2$. We also write down the expression of the flow for one and two channels. 

The fluid starts to flow only above the critical pressure $\Pc=\epsilon_0$ and only along the channel that coincides with the ground state of the directed polymer:
	\begin{equation}
		\label{eqn:flow_groundstate}
		Q_{0,t}(P) = \frac{P-\epsilon_0}{t}
	\end{equation}
	The subscript $0$ indicates that flow is possible only along the ground state. Such a formula holds for $P>\Pc =\epsilon_0$ but less than $P_1$, the pressure at which a second channel opens. 
	
	\subsection{The two channels problem}
	\begin{figure}[h]
		\centering
		\includegraphics[width=0.45\textwidth]{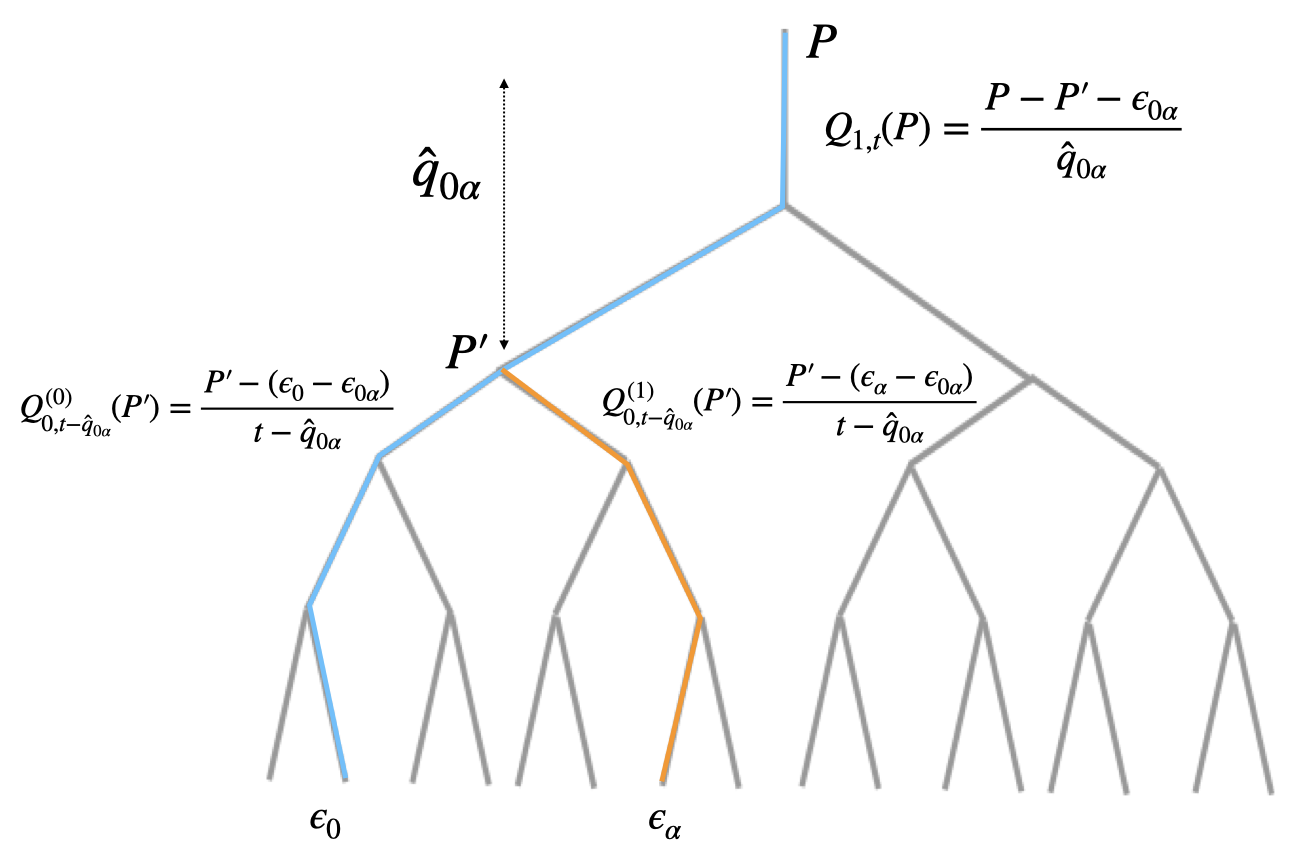}
		\caption{ Schematics for the Cayley tree with two open channels.  }
		\label{fig:jointrees}
	\end{figure}
	Here, flow is possible along two channels (see figure \ref{fig:jointrees}): the ground state with energy $\epsilon_0$, depicted in blue and a second  channel with energy $\epsilon_\alpha$, depicted in orange. These two channels have a common part of length $\qhat_{0\alpha}$; we denote with $\epsilon_{0\alpha}$ the sum of the thresholds along this common portion and $P'$ the pressure at the bottom of the common part, hence  the flow along it reads:
	\begin{equation}
		\label{eq:q_onech_in}
		Q_{1,t}(P) = \frac{P-P'-\epsilon_{0\alpha}}{\qhat_{0\alpha}}
	\end{equation}
	The pressure $P'$ can  be determined using the conservation of the flow:
	\begin{equation}
		\label{eqn:flow_continuity_firstch}
		\frac{P-P'-\epsilon_{0\alpha}}{\qhat_{0\alpha}} = Q_{0, t-\qhat_{0\alpha}}^{(0)}(P') +  Q_{0, t-\qhat_{0\alpha}}^{(1)}(P') 
	\end{equation}
	where $Q_{0, t-\qhat_{0\alpha}}^{(0)}(P')$ is the flow along the subtree containing the ground state, $Q_{0, t-\qhat_{0\alpha}}^{(1)}(P')$ containing the other channel (see the two branches of figure \ref{fig:jointrees}). Since each of these is a single channel of length $t-\qhat_{0\alpha}$, we can use once again Eq.~\eqref{eqn:flow_groundstate}. One has
	\begin{equation}
		Q_{0, t-\qhat_{01}}^{(0)}(P') = \frac{P'-(\epsilon_0-\epsilon_{0\alpha})}{t-\qhat_{0\alpha}} \;, \qquad
		Q_{0, t-\qhat_{0\alpha}}^{(1)}(P') = \frac{P'-(\epsilon_\alpha-\epsilon_{0\alpha})}{t-\qhat_{0\alpha}} 
	\end{equation}
	From equation (\ref{eqn:flow_continuity_firstch}) we derive first the expression for $P'(P)$ and then $Q_{1,t}(P)$:
	\begin{equation}
		P'(P) = \frac{\qhat_{0\alpha}(\epsilon_0+\epsilon_\alpha)+ (t-\qhat_{0\alpha} ) P}{t + \qhat_{0\alpha}} - \epsilon_{0\alpha} \;, \qquad 
		Q_{1,t} (P) = \frac{2}{t+\qhat_{0\alpha}}\left(P-\frac{\epsilon_0+\epsilon_\alpha}{2}\right)
	\end{equation}
	We can now determine the pressure $\tilde{P}_1$ such that  $ Q_{0,t}(\tilde{P}_1) = Q_{1,t}(\tilde{P}_1)$, namely 
	\begin{equation}
		\tilde{P}_1 = \epsilon_0 + \frac{t}{t-\qhat_{0\alpha}} (\epsilon_\alpha-\epsilon_0)
	\end{equation}
	For pressure $P \in (P_0, \tilde{P}_1 )$  we have $ Q_{0,t}(\tilde{P}_1) > Q_{1,t}(\tilde{P}_1)$, this means that the fluid cannot flow in the second channel and the flow rate is given by  $ Q_{0,t}(P)$. Above $\tilde{P}_1$ the second channel is open and the flow rate is given by $Q_{1,t}(\tilde{P}_1)$.	The criterion to select the first excited channel that opens above $\Pc$, is that the pressure $P_1$ is the smallest among all the $\tilde{P}_1$ computed for all possible two-channel geometries. This translates into
	\begin{equation}
		\label{eq:P1min}
		P_1 = \min_{\tilde{P}_1} \tilde{P}_1 =  \epsilon_0 + \min_{\alpha \neq 0} \frac{t}{t-\qhat_{0\alpha}}(\epsilon_\alpha-\epsilon_0)
	\end{equation}
	The channel satisfying the minimum condition is denoted by $\alpha_1$.
	
	\subsection{The three channels problem}
	\label{app:secondch}
	There are three possible configurations for the position of the second excited channel with respect to the ground state and the first one. They each lead to a slightly different expression for the pressure $P_2$, but all simplify to $P_2=\epsilon_{\alpha_2}$ in the limit $t \to \infty$.
	\begin{figure}[h]
		\centering
		\includegraphics[width=0.9\textwidth]{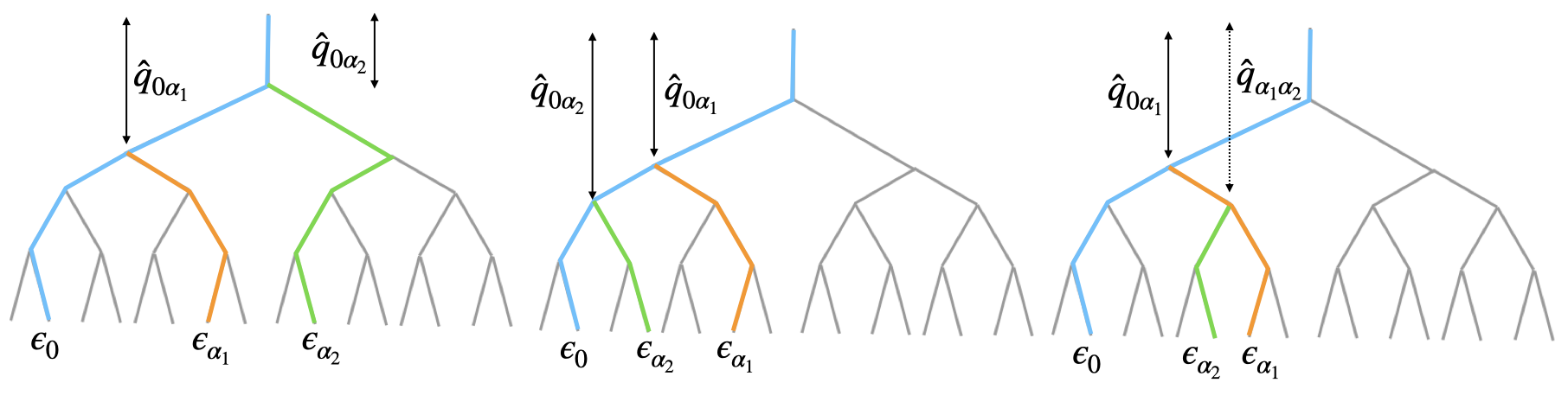} 
		\caption{Schematics for the Cayley tree with three open channels in the three possible geometrical arrangements.}
		\label{fig:p2cases}
	\end{figure}
	
	The first case is the simplest: the second channel opens with a common overlap $\qhat_{0\alpha_2}=\qhat_{\alpha_1\alpha_2}$ with the ground state and the first channel. See figure \ref{fig:p2cases} left. By construction, $\qhat_{0\alpha_2} < \qhat_{0\alpha_1}$. \\
	The pressure $P_2$ reads:
	\begin{equation}
		\label{eqn:pressure2_case1}
		P_2 = \min_{\alpha_2 \neq \{\alpha_1, 0\}} \left[\epsilon_{\alpha_2} - \frac{\qhat_{0\alpha_2}}{t+\qhat_{0\alpha_1}-2\qhat_{0\alpha_2}}(\epsilon_{\alpha_1}+\epsilon_0-2\epsilon_{\alpha_2})\right]
	\end{equation}
	In the limit $t \to \infty$, we saw that $\qhat_{0\alpha_1} = O(1)$; from this and $\qhat_{0\alpha_2} < \qhat_{0\alpha_1}$, it follows that $\qhat_{0\alpha_2}=O(1)$ and $P_2=\epsilon_{\alpha_2}$.
	
	The second case corresponds to the opening of the second excited channel from the ground state with an overlap $\qhat_{0\alpha_2} > \qhat_{\alpha_1\alpha_2} = \qhat_{0\alpha_1}$. See figure \ref{fig:p2cases} middle.
	The pressure $P_2$ reads:
	\begin{equation}
		P_2 = \epsilon_0 - \frac{\qhat_{0\alpha_1}}{t-\qhat_{0\alpha_1}} (\epsilon_{\alpha_1}-\epsilon_0) +  \min_{\alpha_2 \neq \{\alpha_1, 0\}} \frac{t+\qhat_{0\alpha_1}}{t-\qhat_{0\alpha_2}} (\epsilon_{\alpha_2}-\epsilon_0) 
	\end{equation}
	When $t \to \infty$, the previous argument for the first excited channel sets $\qhat_{0\alpha_1}/t \approx 0$. In this limit, the resulting expression for $P_2$ is:
	\begin{equation}
		\label{eqn:pressure2_case2}
		P_2 = \epsilon_0  +  \min_{\alpha_2 \neq \{\alpha_1, 0\}} \frac{t}{t-\qhat_{0
				\alpha_2}} (\epsilon_{\alpha_2}-\epsilon_0) \quad \quad \textrm{when } t \to \infty
	\end{equation}
	This expression is identical to equation \eqref{eq:P1min} with the substitution $\alpha \to \alpha_2$, and applying the same arguments of the first channel we arrive at setting $\qhat_{0\alpha_2}/t \approx 0$, leading $P_2=\epsilon_{\alpha_2}$.
	
	The last case is the mirror of the previous one, with the second channels that opens from the first one with overlap $\qhat_{\alpha_1\alpha_2} > \qhat_{0\alpha_1} = \qhat_{0\alpha_2}$. See figure \ref{fig:p2cases} right. The pressure $P_2$ reads:
	\begin{equation}
		\label{eqn:pressure2_case3}
		P_2 = \epsilon_0 + \frac{t}{t-\qhat_{0\alpha_1}} (\epsilon_{\alpha_1}-\epsilon_0) +  \min_{\alpha_2 \neq \{\alpha_1, 0\}} \frac{t+\qhat_{0\alpha_1}}{t-\qhat_{\alpha_1\alpha_2}} (\epsilon_{\alpha_2}-\epsilon_{\alpha_1}) 
	\end{equation}
	When $t \to \infty$, the previous argument for the first excited channel sets $\qhat_{0\alpha_1}/t \approx 0$. In this limit, the resulting expression for $P_2$ is:
	\begin{equation}
		P_2 = \epsilon_1 + \min_{\alpha_2 \neq \{\alpha_1, 0\}} \frac{t}{t-\qhat_{\alpha_1\alpha_2}} (\epsilon_{\alpha_2}-\epsilon_{\alpha_1}) \quad \quad \textrm{when } t \to \infty
	\end{equation}
	This expression is again similar to equation \eqref{eq:P1min} and applying the same arguments of the first channel we arrive at setting $\qhat_{\alpha_1\alpha_2}/t \approx 0$, leading $P_2=\epsilon_{\alpha_2}$. 
	

\section{B. KPP equation}
	The problem of a directed polymer on the Cayley tree can be studied using the formalism of the discrete KPP equation (see \cite{Derrida1988,Brunet2011} for the original literature). A binary Cayley tree of $t+1$ levels has $2^{t}$ distinct paths denoted by $\alpha$. It can be constructed by considering two independent trees with $t$ levels and joining them together by adding a bond with threshold $\tau$. This hierarchical structure can be exploited when we want to study functions of the energies with a multiplicative form
	\begin{equation}
		G_t(x) = \overline{\prod_{\alpha } g(x - \epsilon_\alpha))}
	\end{equation}
	for some $g(x)$. Indeed $G_{t+1}(x)$ and $G_t(x)$ can be related by the following discrete KPP recursive equation:
	\begin{equation}
		\label{eqn:discrkppdef}
		G_{t+1}(x) = \int d\tau p(\tau) G^2_t(x-\tau)
	\end{equation}
	where $p(\tau)$ is the probability distribution of a single threshold and the initial conditions reads $G_{t=0}(x)=g(x)$.

\begin{figure}[ht]
		\label{fig:kppvelocity}
		\centering
		\includegraphics[width=0.7\textwidth, trim={0 0 0 0},clip]{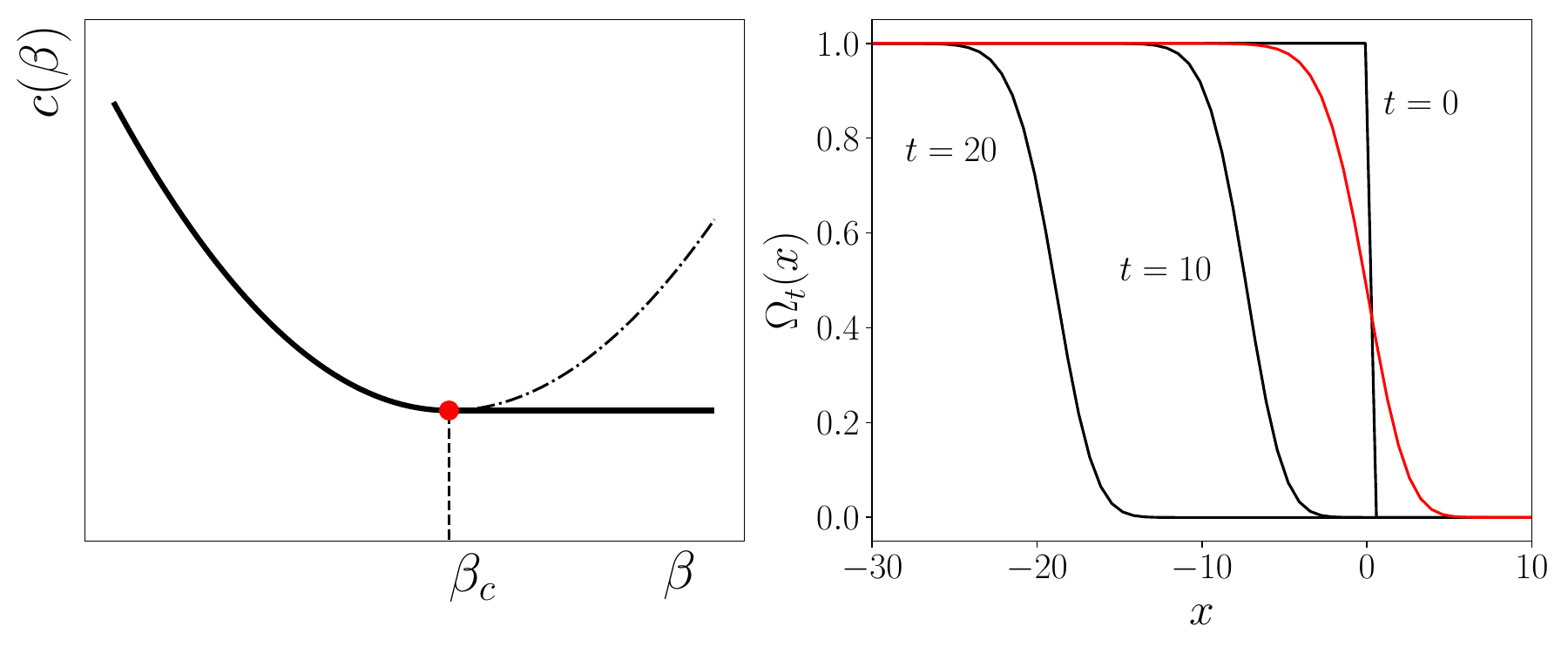}
		\caption{Left panel: the dispersion relation of the velocity (black solid line) as a function of $\beta$. In the Gaussian case, $\tau \sim \mathcal{N}(0,\sigma^2)$, we find from equation \eqref{eqn:cbetaeqn_appendix}: $c(\beta) = \ln 2 /\beta + \beta \sigma^2 /2$ with a minimum at
			$\beta_c = \sqrt{2 \ln 2/}\sigma$.  For $\beta > \beta_c$ the value of the velocity freezes at $c(\beta_c)$. We report in dashed lines the unstable branch of $c(\beta)$ for $\beta > \beta_c$. Right panel: numerical solutions for $\Omega_t(x)$ \eqref{eqn:omegat_multiplicative} for different values of $t$ (black lines). At large $t$, $\Omega_t(x)$ approaches the stationary solution $w_{\rm min}(x)$ (red line), up to a $t$-dependent translation. 
			Note that $w_{\rm min}(x)$ moves exactly at velocity $c(\beta_c)$ while the location of the front $\Omega_t(x)$ is $-c(\beta_c) t$ up to logarithmic corrections \eqref{eqn:P0_appendix}.   }
	\end{figure}

The discrete KPP equation \eqref{eqn:discrkppdef} allows for travelling wave solutions of the form
	\begin{equation}
		G_t(x) = w(x+ct) \qquad \textrm{when} \,\, t \to \infty
	\end{equation}
	where $c$ is the velocity of the traveling wave. When the initial conditions $G_0(x)=g(x)$ satisfy $\lim_{x \to \infty} g(x) = 0$ and $\lim_{x \to -\infty} g(x) = 1$, $c$ is positive and the front moves backward (when the limits are reversed $c$ is negative and the front propagates in the forward direction). The form of the front $w(x)$ satisfies the following fixed point equation:
	\begin{equation}
		\label{eqn:travwavestat_appendix}
		w(x+c) = \int d\tau p(\tau) w^2(x-\tau)
	\end{equation}
	We consider initial conditions of the form $g(x) \stackrel{x \to -\infty}{=} 1 - e^{\beta x}$. By substituting this form into equation \eqref{eqn:travwavestat_appendix} and expanding at first order in $e^{\beta x}$ one obtains an equation for $c$ as a function of $\beta$:
	\begin{equation}
		\label{eqn:cbetaeqn_appendix}
		c(\beta) = \frac{1}{\beta} \log \left( 2 \int d\tau p(\tau) e^{-\beta \tau} \right)
	\end{equation}
	The velocity is shown, as a function of $\beta$, in figure (\ref{fig:kppvelocity}) and presents a minimum at $\beta_c$. It was proven \cite{kolmogorov1937investigation,Bramson83,MAJUMDAR2003161} that the increasing branch of $c(\beta)$ is unstable and that for $\beta > \beta_c$ both the velocity and the front shape freeze, namely $c(\beta > \beta_c) = c(\beta_c)$ and the front shape takes the form $w_{\rm min}(x)$ satisfying the fixed point equation:
	\begin{equation}
		\label{eq:wminrec}
		w_{\rm min}(x+c(\beta_c)) = \int d\tau p(\tau) w_{\rm min}^2(x-\tau)
	\end{equation}
	This equation corresponds to equation (9) of the main text.  Indeed for $g(x)=\vartheta(-x)$, which corresponds to $\beta \to \infty$ We have
 \begin{equation}
			\label{eqn:omegat_multiplicative}
			\Omega_t(x) = \overline{\prod_{\alpha } \vartheta( \epsilon_\alpha-x)} =  \overline{\vartheta(\epsilon_0- x)} \;, \qquad \epsilon_0 = \min_{\alpha} \epsilon_\alpha
		\end{equation}
		where $1-\Omega_t(x)$ is the cumulative distribution of the minimum energy of a directed polymer of $t$ levels. Hence the directed polymer with minimal energy on a Cayley tree with $t$ levels for large $t$ is 
	\begin{equation}
		\label{eqn:P0_appendix}
		P_0 = \epsilon_0 = -c(\beta_c)t + \frac{3}{2 \beta_c} \log t + \varmin 
	\end{equation}
	The extensive part of the minimal energy, $-c(\beta_c) t$, comes from the velocity selection criterion. The subextensive logarithmic correction cannot be obtain by our first order analysis of equation \eqref{eqn:travwavestat_appendix} but was proven by Bramson \cite{Bramson83}. The term $\chi_0$ is a random variable of order $1$ whose cumulative distribution is $1-w_{\rm min}(x)$.

	\subsection{Average number of energy levels above the minimum}
	Using this KPP formalism, we can also study the number of energy levels with an energy less or equal than $x$:
	\begin{equation}
		\label{eqn:numstatesabsolute}
		n_t(x) = \sum_{\alpha} \vartheta(x-\epsilon_\alpha)
	\end{equation}
	Indeed by choosing as initial condition $g(x) = 1 + \vartheta(x) (\lambda-1)$ we can obtain the generating function of $n_t(x)$, namely $\Psi_t(x;\lambda) = \overline{\lambda^{n_t(x)}}$.
	However, we are interested in the average number of energy levels having an energy $\epsilon_\alpha$ bigger than the minimal energy $\epsilon_0$ of an amount $x$. We denote this quantity by:
	\begin{equation}
		m^{(\rm full)}_t(x) = \overline{ \sum_{\alpha} \vartheta(x+\epsilon_0-\epsilon_\alpha)} = \overline{n_t(\epsilon_0 + x)}
	\end{equation}
	This quantity is more complex than $n_t(x)$ since it involves the minimal energy $\epsilon_0$. However, following an approach introduced in \cite{Brunet2011}, it is possible to derive an equation for $m^{(\rm full)}_t(x) $ that we can solve numerically.
	We sketch here the derivation. The reader uninterested to the technical details can skip this section. 
	
	The first step is to introduce the generating function of $m_t^{(\rm full)}(x)$ with $|\lambda|<1$:
	\begin{equation}
		\label{eq:mdef}
		\chi_t(x; \lambda) = \overline{\lambda^{m_t^{(\rm full)}(x)}}
	\end{equation}
	This function satisfies the following identity:
	\begin{equation}
		\label{eqn:chitidentity_appendix}
		\chi_t(x; \lambda) = \int dx' \; \overline{\lambda^{n_t(x' + x)} \delta(x' - \epsilon_0)} =1 + \partial_x \int dx' \; 
		\overline{\lambda^{n_t(x' + x)} \vartheta(\epsilon_0 - x')  }
	\end{equation}
	To prove the validity of \eqref{eqn:chitidentity_appendix} we used $(\partial_x \lambda^n )\vartheta = (\partial_{x'} \lambda^n )\vartheta = \partial_{x'}( \lambda^n \vartheta) - \lambda^n (\partial_{x'} \vartheta)$.
	The integrand of equation \eqref{eqn:chitidentity_appendix} satisfies a discrete KPP equation \eqref{eqn:discrkppdef} since it can be recast into a multiplicative form:
	\begin{equation}
		\lambda^{n_t(x' + x)} \vartheta(\epsilon_0 - x')   = \prod_{\alpha } \lambda^{\vartheta(x' + x - \epsilon_\alpha)} \vartheta(\epsilon_\alpha - x') 
	\end{equation}
	In this case, the initial condition reads $g(x';x,\lambda) = \lambda^{\vartheta(x' + x)} \vartheta(-x')$.
	The average number of energy levels above the minimum can be written as:
	\begin{equation}
		\overline{m^{(\rm full)}_t(x)} = \left. \partial_{\lambda} \chi_t(x; \lambda) \right |_{\lambda = 1}
	\end{equation}
	We thus expand for $\lambda = 1 - \varepsilon$
	\begin{equation}
		\label{eqn:integrandexp_appendix}
		\overline{\lambda^{n_t(x' + x)} \vartheta(\epsilon_0 - x') } = 
		\Omega_t(x') - \varepsilon R_t(x'; x) + O(\epsilon^2) 
	\end{equation}
	The function $\Omega_t(x)$ is defined in \eqref{eqn:omegat_multiplicative}.
	This expansion leads to:
	\begin{equation}
		\label{eq:mviar_app}
		\overline{m^{(\rm full)}_t(x)} = \int dx' \; \partial_z R_t(x'; x) = \int dx' \; r_t(x'; x)
	\end{equation}
	where we set $r_t(x'; x) = \partial_x R_t(x'; x)$. 
	Now, by plugging the expansion \eqref{eqn:integrandexp_appendix} in \eqref{eqn:discrkppdef} and taking the first order in $\varepsilon$, we obtain a linear equation for $R_t(x'; x)$ 
	\begin{equation}
		\label{eq:Rlin}
		R_{t+1}(x'; x) = 2 \int d\tau \; p(\tau) \Omega_t(x' - \tau) R_t(x' - \tau; x) \;, \qquad R_{t=0}(x'; x) = \frac 1 2\vartheta(x' + x)\vartheta(-x') \;.
	\end{equation}
	Moreover because of the linearity of \eqref{eq:Rlin}, $r_t(x; z)$ satisfies 
	\begin{equation}
		\label{eq:rlin_app}
		r_{t+1}(x'; x) = 2 \int d\tau \; p(\tau) \Omega_t(x' - \tau) r_t(x' - \tau; x) \;, \qquad r_{t=0}(x'; x) = \frac 1 2 \delta(x' + x)
	\end{equation}
	By numerically solving simultaneously the equation for $\Omega_t(x)$ and \eqref{eq:rlin_app}, one has direct access to the $t$ dependence of $\overline{m_t^{(\rm full)}(x)}$ \eqref{eq:mviar_app}.
 \newline
 As reported in the main text, the same equations hold for the determination of $m_{\qhat}(x)$. Indeed equation \eqref{eq:rlin_app} changes as:
 \begin{equation}
		\label{eq:rqlin_app}
		r_{\qhat+1}(x'; x) = 2 \int d\tau \; p(\tau) w_{\rm min}(x' - \tau + c(\beta_c)) r_{\qhat}(x' - \tau; x) \;, \qquad r_{\qhat=0}(x'; x) = - w'_{\rm min}(x'+x)
	\end{equation}
 Yielding:
 \begin{equation}
		\label{eq:mqr}
		m_{\qhat}(x) = \int dx' \, r_{\qhat}(x'; x) \;.
	\end{equation}

  \subsection{Numerical integration of discrete KPP equation}
	The directed numerical integration of $\Omega_t(x)$ \eqref{eqn:omegat_multiplicative} is not feasible at large $t$ as the position of the front moves in the backward direction. In order to maintain the integration limits fixed in a window $[x_{\rm min}, x_{\rm max}]$, one needs to shift the front by fixing its position as $t$ grows. Moreover to ensure a better numerical stability we evolve  $H_t(x)=e^{-\beta_c x} (1-\Omega_t(x))$ then the equation for \eqref{eqn:omegat_multiplicative} becomes:
	\begin{equation}
		H_{t+1}(x) = \int d\tau \tilde{p}(\tau) [2H_t(x-\tau)-e^{\beta_c(x-\tau)}H_t(x-\tau)]
	\end{equation}
	with $\tilde{p}(\tau)=p(\tau) e^{-\beta_c \tau}$ and initial condition $H_0(x)=\vartheta(x) e^{-\beta_c x}$. To fix the position of $H_t(x)$ we compute numerically (Riemann sum) $b_t=\int dx H_t(x) e^{\beta_c x}$. At each step $t$ if $b_t-b_0>0$ we make the substitution $H_t(x) \to H_t(x-(b_t-b_0))$. As we shift we should assign a value of $H_t(x)$ to the points in $[x_{\rm min}, x_{\rm min}+b_t-b_0]$; to do so interpolate the last $10$ points using a spline.
	We employed $p(\tau)=\frac{1}{\sqrt{2\pi}} e^{-x^2/2}$ and discretized our integration limits $[x_{\rm min}, x_{\rm max}]$ in $N=150000$ points with  $x_{\rm min}=-10000$ and $x_{\rm max}=30$. The equation for $r_t(x; z)$ \eqref{eq:rlin_app} is solved in parallel to $H_t(x)$ taking into account the aforementioned translations. We use the same procedure to solve for $r_{\qhat} (x; z)$.

 \subsection{Large $\hat{q}$ limit of maximal overlap energy levels }
	In Eq.~\eqref{eq:mqlim} of the main text, we report the convergence of $\overline{m_{\qhat}(x)}$ to $e^{\beta_c x}$ as $\qhat \to \infty$ . To get further insights about the large $\qhat$ behavior of $\overline{m_{\qhat}(x)}$, we further analyze Eq.~\eqref{eq:rqlin_app}. As it is obtained from the linearization of the full KPP equation \eqref{eqn:discrkppdef} which supports travelling waves, it still induces a translation at velocity $c(\beta_c)$. In order to get rid of this translation effect, we introduce the shifted quantity $r_{\qhat}^{(s)}(x'; x) = r_{\qhat}(x' - c(\beta_c) \qhat; x)$. In this way, Eq.~\eqref{eq:rqlin_app} becomes
	\begin{equation}
		r_{\qhat+1}^{(s)}(x'; x) = 2 \int d\tau \; p(\tau) 
		w_{\rm min}(x' - \tau - c(\beta_c)) 
		r_{\qhat}^{(s)}(x' - \tau - c(\beta_c); x) = 
            \int dx'' \mathcal{L}(x', x'') r_{\qhat}^{(s)}(x''; x)
	\end{equation}
	where in the last equality we changed the integration variable setting $x'' = x' - \tau - c(\beta_c)$ and implicitly defined the linear operator $\mathcal{L}$. Thus, one can pass from $\hat q \to \hat q + 1$ by applying a fixed matrix $\mathcal{L}(x', x'')$\footnote{We use the expression \emph{matrix}, although compact linear operator would be more appropriate in the mathematical jargon.}.  Since the matrix $\mathcal{L}$ is independent of $\hat q$, iterating this equation $\hat q$ times, we can express  $r_{\hat q}^{(s)}$ in terms of the initial condition at $\hat q = 0$
 \begin{equation}
     r_{\qhat}^{(s)}(x'; x) = \int dx'' \mathcal{L}^{\hat q}(x', x'') r_{0}^{(s)}(x''; x) =  - \int dx'' \mathcal{L}^{\hat q}(x', x'') w'_{\rm min}(x'' + x)
 \end{equation}
 where we denoted as $\mathcal{L}^{\hat q}$ the $\hat q$-th matrix power of $\mathcal{L}$ and in the last equality we used \eqref{eq:rqlin_app}.
 In the limit $\qhat \to \infty$, $\mathcal{L}$ acts as a projector on the eigenspace with largest eigenvalues. Taking the derivative of Eq.~\eqref{eq:wminrec} with respect to $x'$, we obtain
	\begin{equation}
		\label{eq:wmineigen}
		w_{\rm min}'(x') =  \int dx'' \mathcal{L}(x', x'') w'_{\rm min}(x'') \;.
	\end{equation}
 which shows that $w_{\rm min}'(x')$ is an eigenvector with eigenvalue $1$. 
 We also have $-w_{\rm min}'(x')>0$, as it corresponds to the probability density of the minimum.  Since $\mathcal{L}(x', x'')>0$, Perron-Frobenius theorem ensures that it corresponds to the largest eigenvalues which equals $1$. Therefore, without further knowledge of the full spectrum of $\mathcal{L}$, we know that the remaining eigenvectors lie inside the unit circle. We can thus expand 
 the initial condition on the eigenvectors as
 \begin{equation}
 \label{eq:rexpansion}
     r_0^{(s)}(x'; x) \equiv -w'_{\rm min}(x' + x) = - A(x) w'_{\rm min}(x') + R(x'; x)
 \end{equation}
 where the first term contains the components on the leading eigenvector and $R(x'; x)$ stays for the remainder. The repeated application of $\mathcal{L}$ projects on $w'_{\rm min}(x') $ and thus
\begin{equation}
		\lim_{\hat{q} \to \infty} r_{\qhat}^{(s)}(x'; x) = - A(x) w'_{\rm min}(x') 
  \quad \Rightarrow\quad
  \lim_{\hat{q} \to \infty} m_{\hat{q}}(x) = - A(x) \int dx' w'_{\rm min}(x')
  = A(x)
\end{equation}
where we used \eqref{eq:mqr} and in the last equality we performed the integration over $x'$ using that $-w'_{\rm min}(x')$ is a normalised probability density. Thus, the large$-\hat{q}$ asymptotic behavior $m_{\hat{q}}(x)$ is given by the coefficient $A(x)$ in \eqref{eq:rexpansion}. However, since the operator $\mathcal{L}$ is not self-adjoint, 
in order to compute the projector on the eigenvector in \eqref{eq:wmineigen}, we need to determine the left eigenvector $\ell_{\rm min}(x')$ corresponding to $w'_{\rm min}(x')$, which satisfies
\begin{equation}
    \int dx' \ell_{\rm min}(x') R(x'; x) = 0 \;, \qquad \int dx' \ell_{\rm min}(x') w_{\rm min}'(x') \neq 0 \;.
\end{equation}
so that we can express the coefficient $A(x)$ as
\begin{equation}
\label{eq:mqlim_app}
    A(x) = \frac{\int dx' \ell_{\rm min}(x') w_{\rm min}'(x' + x) }{\int dx' \ell_{\rm min}(x') w_{\rm min}'(x') }
\end{equation}
Computing the adjoint of $\mathcal{L}$, we see that $\ell_{\rm min}(x')$ must satisfy
\begin{equation}
\label{eq:lmin}
		\ell_{\rm min}(x') = 
		2 w_{\rm min}(x') \int dx'' \; \ell_{\rm min}(x'') p(x'' -x' - c(\beta_c)) 
\end{equation}
%
%
	The explicit form of $\ell_{\rm min}(x)$ cannot be determined in general as it depends on the specific form of the threshold distribution $p(\tau)$. So, it might look surprising that eventually the large $\hat{q}$ drastically simplify to a universal form, but a subtle mechanism is at play. Indeed, in the limit $x' \to -\infty$, we assume $\ell_{\rm min}(x') \sim e^{-\tbeta x'}$. Plugging it in Eq.~\eqref{eq:lmin} and using that $w_{\rm min}(x \to -\infty) = 1$, we can determine the value of $\tbeta$. We obtain
	\begin{equation}
		2 \int dx' \; e^{-\tbeta x'} p(x' -x - c(\beta_c)) = e^{-\tbeta x}
	\end{equation}
	and after the change of variables $\tau = x' - x - c(\beta_c)$, it coincides with Eq.~\eqref{eqn:cbetaeqn_appendix} with $\tbeta = \beta_c$. 
 Since $w_{\rm min}'(x') \stackrel{x' \to -\infty}{\longrightarrow} x' e^{\beta_c x'}$, we see that both the numerator and denominator in Eq.~\eqref{eq:mqlim} are formally infinite as the integrand diverges for $x' \to -\infty$. So the ratio in Eq.~\eqref{eq:mqlim_app} needs to be evaluated by a limiting procedure. In order to regularize we introduce a cutoff $\Lambda$ on the left tail, replacing $\ell_{\rm min}(x) \to \vartheta(x + \Lambda)\ell_{\rm min}(x)$, so that the result should be recovered in the limit $\Lambda \to \infty$. Then, we have
	\begin{equation}
		\lim_{\hat{q} \to \infty} m_{\hat{q}}(x) \equiv A(x) 
		= \lim_{\Lambda \to \infty} \frac{\int_{-\Lambda}^\infty dx' \; \ell_{\rm min}(x') w'_{\rm min}(x + x')}{\int_{-\Lambda}^\infty dx' \; \ell_{\rm min}(x') w'_{\rm min}(x')} 
		=
		\lim_{\Lambda \to \infty} \frac{e^{\beta_c x} \int_{-\Lambda}^0 dx \; (x' + x) + O(1)}{\int_{-\Lambda}^0 dx' \; x' + O(1)} = e^{\beta_c x}
	\end{equation}
	which gives the expected exponential behavior.

  \section{C. Determination of the flow}
   
  \subsection{Large Pressure limit}
  The hierarchical structure of the tree allows to write a recursive equation in $t$ for the flow:
 \begin{flalign}
 & Q_t(P) = (P-P'-\tau)_+ \\
 &  (P-P'-\tau)_+ = Q_{t-1}^{(1)} (P') + Q_{t-1}^{(2)} (P')
 \end{flalign}
 the superscripts above indicates two different branches of the tree. The second equation come from Kirchhoff law applied at the node joining a new tube with two different sub-trees. This node is at pressure $P'$ which must be found self-consistently as a function of $P$. As reported in the main text, this is a non-feasible task for generic $t$. However in the $P \to \infty$ regime all the channels are open, the flow takes a simple linear form:
 \begin{equation}
     Q_t(P) =\kappa_t(P-P_t^{*})
 \end{equation}
 Inserting this ansatz in the second equation we get:
 \begin{equation}
     P' = \frac{P-\tau+\kappa_{t-1}(P_{t,1}^*+P_{t,2}^*)}{1+2\kappa_{t-1}}
 \end{equation}
A recursive equation for the  permeability,
\begin{equation}
    \kappa_t = \frac{2\kappa_{t-1}}{1+2\kappa_{t-1}}
\end{equation}
 together with $\kappa_1=1$ allows to find the solution :
\begin{equation}
    \kappa_t = \frac{2^{t-1}}{2^t - 1} 
\end{equation}

For large trees $\kappa_t \approx 1/2$. As we found $\kappa_t$, we can find the  formula for $P_t^*$:
\begin{flalign}
    & P_t^* = \tau + \frac{1}{2} (P_{t, 1}^* + P_{t, 2}^*) 
\end{flalign}
As all the $P_{t,1}^*$ and $P_{t,2}^*$ are independent random variables, coming from separate sub-trees, the average $P_t^*$ reads:
\begin{equation}
    \overline{P_t^*} = \overline{\tau} + \overline{P_{t-1}^*} = t \overline{\tau}
\end{equation}

 \subsection{Exact numerical solution of the flow for the Cayley tree}
	The Darcy flow on the Cayley tree can be exactly numerically solved for moderate $t$. 
	\label{app:darcynumsolution}
	\begin{figure}[h]
		\centering
		\includegraphics[width=0.5\textwidth]{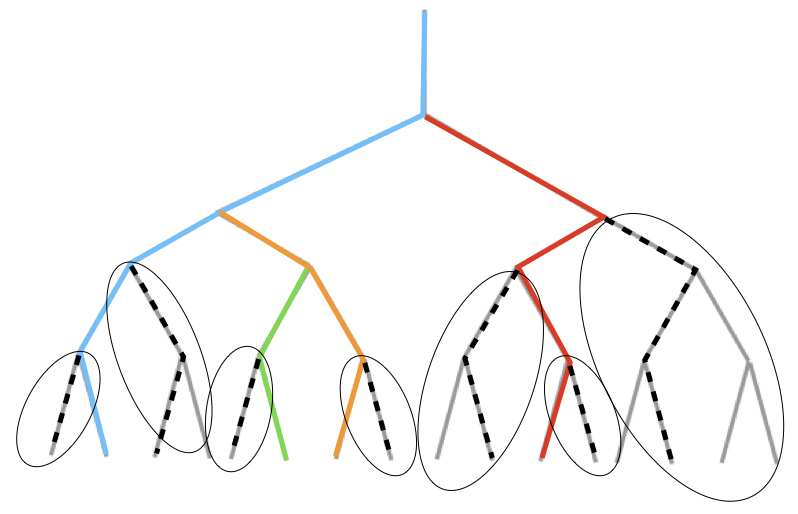}
		\caption{Sketch of the algorithm used for the exact solution of the flow in the Cayley tree. }
		\label{fig:optim_scheme}
	\end{figure}
	The solution consists in finding the pressures $\Pc, P_1, P_2, \dots$ at which each new channel opens. The algorithm we use to produce the Fig. 3 middle and the Fig. 3 right of the main text is a simplified version of the general algorithm valid for any directed network and discussed in the supplementary material of \cite{ChenDeLucaRossoTalon}. In practice, the critical pressure $\Pc$ is the ground state of the associated directed polymer model. The other pressures are found following an iterative procedure. At each step, we denote by $\mathcal{C}_s$ the subtree made of open channels exclusively for $P \in [P_s, P_{s+1}]$ (where $P_{s+1}$ is not yet determined). Then,
	\begin{enumerate}
		\item Using Kirchhoff's equations, we determine the pressure at each node of $\mathcal{C}_s$ as a function of the applied pressure $P$ (these functions are linear in $P$). 
		\item We find all the subtrees connected to a node in $\mathcal{C}_s$ which are still closed. In figure \ref{fig:optim_scheme} they are highlighted by ovals.
		\item For each subtree, we find its ground state, denoted by $\mathcal{E}_s^{(1)}, \mathcal{E}_s^{(2)}, \dots$. They are indicated with dashed lines in figure \ref{fig:optim_scheme}.
		\item We find the minimal applied pressure $\tilde{P}_{s}^{(1)}, \tilde{P}_{s}^{(2)}, \dots$ at which the ground state of each subtree opens. This is done by setting the node pressure at the root of each subtree (determined in step (1) using Kirchhoff's equation) equal to $\mathcal{E}_s^{(1)}, \mathcal{E}_s^{(2)}, \dots$. 
		\item We pick as a new channel the one with minimal opening pressure, namely $P_s = \min_\alpha \tilde{P}_s^{(\alpha)}$.
	\end{enumerate}
	In numerical simulations, we stop at $t=23$ due to the exponential growth in $t$ of the number of configurations on the Cayley tree.
	\label{app:chnexcstates}
	
\end{document}